# A case for intranet-based online portal for undergraduate Computer Science teaching
(An approach for computer-assisted-learning)

K. V. Iyer
Dept. of Computer Science & Engg.
National Institute of Technology
Tiruchirapalli 620015, India
kvi@nitt.edu

*Abstract*—Our proposal for selective subjects especially those involving intensive problem-solving assignments and/or tutorials, such as Introduction to Algorithms and Data structures, Discrete Mathematics, Coding Theory, Number theory, Combinatorics and Graph Theory (CGT), Automata theory, is to supplement lectures with a moderated online forum against an intranet portal. By way of illustration we take the example of a restricted view of OEIS (http://oeis.org). The restriction can be w.r.t. sequences in OEIS that are directly relevant to say CGT. N.J.A.Sloane's OEIS is a collection of over 2,39,147 integer sequences and their properties. In particular OEIS contains definitions of many combinatorial structures, dense range of interpretations, generating functions and conjectured ones, cross references within OEIS and to outside resources, references to texts and technical articles, codes in Maple, Mathematica etc. For organizing courses such as the above mentioned, a first task is to partially create an OEIS-like instructor-moderated portal in a university intranet. During the course of lectures and tutorials students are invited to contribute to the portal and these may be augmented/approved by instructors suitably, to find a place in the portal. By this many concepts can be conveyed to the students in an interesting way with the desired results. In the arguments presented, examples related to Combinatorics and Graph Theory are given.

I. INTRODUCTION

Consequent to the availability of the Internet and hardware systems, web-based computer education, popular in many educational institutions depends on the availability of high quality learning resources such as electronic books [1] and technical articles, multimedia enabled presentations, refereed technical articles etc. In a related context web-based problem-solving is a well-known learning activity where, among other things, students are required to use a search-engine to extract contents from the web, pertaining to a target theme and summarize and classify the content (it is known that the task of abstracting and classifying page-ranked web-search results can be efficiently automated). A major goal in these activities is to enrich the learning experience of students as is the case with intelligent tutoring systems. It is argued that web-based systems for learning have these advantages [2]: long-distance learning, being a complement for traditional teaching and as a support for life-long learning. Conventional intranet-based education relies on a server onto which lecture notes, handouts, hyperlinks to selected Wiki pages, assignments and solutions etc. are preloaded by the instructor. To some extent these can cope-up with the rapid changes in Computer Science as a result of technological advances. However these approaches lack the feature of interactiveness in computer-based problem-solving. The generic framework ActiveMath described in [2] is web-based, adaptive and interactive.

Techniques in computer-aided-instruction-oriented systems are known to be teacher-assisted or otherwise. These systems often incorporate online assessments of students such as a quiz. On the negative side, use of such systems are known to create disengagement in the users' minds. As mentioned in [3] the different forms of disengagement from the task at hand include engaging in off-task behavior such as net-surfing, exploiting system's features in a quiz such as repeated guessing, working carelessly and using the system in ways unrelated to the stated work. It has been pointed out that boredom and confusion precede such behaviors although such computer-based learning systems are seen to be quite engaging in general and are useful when the number of students in a class is large.

The suggestion described herein is a student-centric use of a university intranet-web portal associated with a course that dynamically evolves with the progress of the course – this approach of describing and expanding knowledge, developing comprehension, analytic ability and inquisitiveness in a subject can be classified under computer-assisted learning in some sense. The approach outlined herein can be regarded as a positive step in enhancing a learning environment ("learning through examples") for undergraduate Computer Science teaching. For selective subjects (those involving intensive problem-solving tutorials and exercises), such as Introduction to Algorithms and Data structures, Discrete Mathematics, Coding theory, Number theory, Combinatorics and Graph Theory (CGT), Automata theory, our proposal is to supplement lectures with a teacher-moderated online forum on a university intranet portal with restricted access to a class of students. In this the role of a teacher and his/her team is augmented by the role of students and it promotes collaborative learning as well, in computer-free classrooms. It also supports the future classroom models where students will listen to lectures at their own convenience in his/her environment and will get engaged in group activities in classrooms.

By way of illustration we take the example of a restricted view of the Online Encyclopaedia of Integer Sequences (OEIS: see [4]). The restriction can be w.r.t. sequences in the OEIS that are directly relevant to say CGT. The rest of the paper briefly mentions about OEIS and then about the ways and means a similar portal can be developed in a university intranet for subjects related to Computer Science and perhaps Mathematics for undergraduate teaching. For the purpose of illustrations basic concepts in CGT is required (see for example, [5,6,7]).

## II. THE OEIS AS A LEARNING RESOURCE

First, a few highlights about the OEIS are given. The proposition of an intranet portal for a specific course may be viewed as a miniature form of OEIS although with many differences in content as well as in the method of construction and usage.

Founded in 1964 by N.J.A.Sloane, OEIS, the now-online learning resource is a collection of over 2,39,147 integer sequences and their properties. In particular, as of now the OEIS contains definitions of many combinatorial structures, dense range of their interpretations, generating functions and conjectured ones, cross references within OEIS and to outside resources, references to texts and technical articles, codes in Maple, Mathematica etc. OEIS incorporates a search feature allowing keyword-based search spanning the entire contents. There are other features as well to use the OEIS. Entries and modifications to the OEIS contributed by integer sequences enthusiasts across the world are reviewed and then approved by a team of subject experts. The overall contents of OEIS is to be viewed at a research level. A motivated student can explore the OEIS, fairly systematically and enhance his/her knowledge in Discrete Mathematics with additional learning support. The popularity of OEIS is evident from the numerous citations it has received in serious mathematical literature. A typical entry identified by an OEIS sequence number (e.g., A136328 in this case) looks like the one given in Illustration 1 (modified to remove hyperlinks). This sequence lists the first sixteen values of the Wiener index of Odd graphs. As a prerequisite, this page requires notions such as graphs, sets and related operations, univariate polynomials, cycles in graphs, induced subgraphs permutations of $[n]$, oddness, Wiener index, distance in graphs, computer algebra systems. At present OEIS is managed by the OEIS foundation.

## III. OEIS-LIKE INTRANET-PORTAL-BASED COMPUTER SCIENCE TEACHING

For organizing courses such as the above mentioned, the first task is to create partially, an OEIS-like portal in a university intranet on a client-server architecture. The portal can be built using *XML* technology and can be made accessible via standard browsers. The initial entries are due to an interest group with an identified moderating team. During the course of lectures and tutorials, students should be invited to contribute to the portal and these may be augmented/approved by teachers and graduate students to find a place in the portal. Submissions to the portal can be addressed to a designated mailbox monitored by the moderators. Contributions from the instructors are aimed to be at first year graduate level so as to motivate the undergraduates. A calendar of associated tasks drawn by faculty member will set the timelines. Some of the activities such as supplying figures, formatting, suggesting computer codes, cross referencing pointers, verifications, pointers to other resources can be assigned to a skilled groups of students. By this, many concepts in a specific subject can be conveyed to the students in an interesting way with the desired results. An added advantage is that assignments and reading exercises can be against the portal. This is beneficial when the class strength is large, say 100 which is generally the case in foundational courses. We emphasize in later discussions the dynamic nature – in terms of growth and restructuring - of the portal-idea: For example in the OEIS-context addition suggestions to A136328 in Illustration 1 could be: (a) Under COMMENTS: Properties of cycles in $O_n$. (b) Under REFERENCES: Pointer to a proof that $O_n$ is distance regular. (c) Under LINKS: A table of Hosoya-Wiener polynomials of $O_n$ for different $n$ values. An exceptional student/teacher may like to add a new page to the OEIS on the Wiener index of line graphs of odd graphs.

### A. An A-CGT-portal and its construction

As a detailed example, a portal on *Special Graphs and Combinatorial Objects* is appropriate to be associated with a first-level course on CGT (in the sequel this is referred to as A-CGT-portal). Graphs such as $K_n$, $K_{m,n}$, $C_n$, $S_n$, *Ladder graphs, binary hypercubes, Fibonacci trees, Fibonacci weighted trees* etc., will serve as starting candidate objects. An A-CGT-portal will contain in each logical page, primary attributes such as *definition(s), figures, constructions and algorithms to generate the objects, properties, related objects/graphs, a more-to-explore set of references/reading materials, historical notes/significance, and remarks by students/instructors*. The properties and references will strive to capture a body of knowledge appropriate to CGT *per se* and to other courses requiring CGT as a prerequisite. Typical sources of knowledge include student submissions of assignment problems, copyright unprotected and free web/other resources, sources for which explicit permissions have been obtained, materials developed by instructors. Associated with an A-CGT-portal will be a corpus of prerequisite keywords and phrases – e.g., *graphs,*

*combinatorial structures, graphs characterized by one or more parameters, induced subgraphs, isomorphism, trees, power set, permutations and combinations, Pascal's triangle, partitions, recurrence relations, generating functions* etc. A backward link will be associated with each logical page of an A-CGT-portal – the link will be from a perquisite box(es) built from the corpus to the relevant logical page(s). It is possible to recognize the following two types of prerequisite boxes:

*Type P1:* This comprises those (*lightweight*) keywords/phrases that are acquired in one or two sittings by a student. These are covered by 1-2 lecture equivalents or textbook chapters.

*Type P2:* This encompasses (*loaded*) keywords or phrases each of which point to 1-4 short write-ups or pages of references e.g., planarity and embeddings, perfect graphs.

A more simple approach is the logical page attribute *allowable perquisite courses*. A *color encoding* of a logical page's background or of the primary attributes will indicate which A-CGT-portal pages are part of the CGT course and which ones are considered outside the purview of the course. Alternatively and/or as an added feature a *degree of relevance* indicator computed as a weighted average based on the prerequisite boxes for each logical page of an A-CGT-portal is considered as a logical page attribute. The relevance will be to lecture notes and textbook chapters and sections.

The demarcation between the two types P1 and P2 is rather fuzzy in the case of many key terms and it depends on the knowledge and comprehension already acquired by the students. In general each keyword or phrase, via a hyperlink is made to point to a short page containing relevant definitions and/or concepts, illustrated if appropriate. In certain instances the page pointed at will be already part of the A-CGT-portal.

A typical logical page in an A-CGT-portal will cover a special graph or it can describe a graph class. We illustrate a logical page with the example of a *Wheel graph* . By definition $W_n = C_{n-1} + v$, where $v$ is a new vertex connected to all vertices in $C_{n-1}$. The construction is implicit in the definition. A reference will be to a Wiki-source page. Some of the properties include:

(a) For large $n$, the ratio number of edges to number of edges approaches 2.
(b) $W_n$ is planar with a unique embedding.
(c) $W_n$ is a self-dual graph.
(d) $W_n$ is a *Halin graph*.
(e) $W_n$ is a (triangulated) perfect graph.

Property (d) is color coded to inform the students that topics in perfect graphs are part of a second level course. Accordingly we build a prerequisite box. With queries from students, an instructor will provide the definition of perfect graph in an assignment with examples and ask for a proof that $W_n$ is perfect. An addition under *related graphs* will be the *Gear graph* which is a wheel with a vertex added between each adjacent pair of vertices in the outer cycle of the wheel. This is likely to be developed by students and will evolve into a separate logical page of an A-CGT-portal. Although this may enhance the scope of the topics in CGT it will be viewed positively by many students. As opposed to a special graph a logical page may describe *k*-regular graphs, binary trees, binomial trees, spanning trees of complete graphs, *R-L-C* electrical networks, Flow networks, Reliability networks etc. For an electrical sciences major *R-L-C* networks is a reading assignment with type P2 prerequisite box comprising phrases such as *active networks and Ohm's law, Kirchhoff's current/voltage laws, network theorems, dual networks, Cut-sets and Tie-sets.* For Computer Science majors recommended readings will be binomial trees and Flow networks pages. It is easy to see that with the progress of a CGT course, an A-CGT-portal will evolve, with learning benefits to the students in general.

*B. A-CGT portal's influence on teaching methodology and on students*

It is expected that enthusiastic teachers will always offer challenges via an A-CGT-portal. Teacher's challenges in the form of assignments and tests and open questions and students' responses are analysed and best understood against the class average grade. Assuming that the possible student assessment grades are A, B, C, D, E, F it is possible to understand the level of knowledge and perseverance of the students as given below. Let $T$ denote the term GPA of a student so far in all subjects relevant to CGT, cleared by a student. We group the students based on $T$-value-mapped grades and list a few descriptive attributes and conclusions for each group.

*Group 1*: $T \varepsilon$ {A, B}
---Highly motivated in CGT, Able to solve a complete problem set by oneself, Able to pose well-formed problems, Self-initiated to use learning resources.
Conclusion: Will be able to contribute entries to A-CGT-portal e.g., algorithms for generating/drawing graphs.
*Group 2*: $T \varepsilon$ {C, D}
---Fairly motivated in CGT, Able to understand the subject matter well with some effort, Completes exercises with rewards, Uses learning resources in a goal-oriented manner.
Conclusion: Will find A-CGT-portal as a learning resource and with guidance can contribute to page attributes.
*Group 3*: $T \varepsilon$ {E, F}
---Poorly motivated in CGT, Finds difficulty in getting started with the problems, Relies much on group activities, Needs significant effort to cross the subject.
Conclusion: Will be able to use the prerequisite boxes and will learn from A-CGT-portal at a slow pace.

The types of problems posed/arising through a typical A-CGT-portal page are as mentioned below.

(a) Easily solvable by 60% of the students.
(b) Solvable with a knowledge of higher Mathematics developed through the CGT course and prerequisites.
(c) Solvable via a programming language such as *Java/C++*.
(d) Conjectures for which solution approaches are sought.

(e) Problems and solutions from international contests such as Math Olympiad, ACM/ICPC.

The exercises in tests or in assignments in the context of an A-CGT-portal will test the following:

(i) Understanding of the definitions and key terms, properties etc. listed in an A-CGT-portal page.
(ii) Understanding of notions and concepts via problems not immediately apparent from an A-CGT-portal page.

For Group 2/3 students hints and suggestions in a tutorial session include:

(a) Clarifications via explanations (maybe in a native tongue) explaining associated concepts/examples and/or via given assignments/quizzes explicitly establishing the connections.
(b) Supplementary reading in CGT and suggestions to explore specific A-CGT-portal pages and prerequisite boxes.

In general these will require additional reading of A-CGT-portal pages. For Group 3 students counseling is also appropriate e.g., an advice to learn at a slower pace such as learning in a group under a supervision. Based on the effective usage of an A-CGT-portal and contributions to it, a teacher may expect that a Group 3/2 student will slide into Group 2/1 with the progress of a CGT course.

The number of the different types ((a) through (e) mentioned above) of exercises in assignments and tests will be decided by the percentage of students in the different Groups. We illustrate motivational quiz problems with the following two examples:

Ex. 1. A question in CGT asks to find the number of vertices and edges in the graph *block_n* defined thus: A basic block is a graph $G_r = (V_r, E_r)$; $V_r = (r, A, B, C, D, E, F)$ and $E_r = (rA, rB, AC, AD, BE, BF, CD, CE, DF, EF)$. A *block_1* is built from two copies of the basic blocks rooted at *r_1* and *r_2* with the added edge *r1r2*. A *modified block_1* is a *block_1* wherein the edge *r1r2* is replaced by the path *r1—x—r2* where *x* is a newly added vertex. A *block_2* consists of two copies of the *modified block_1* having the new vertices *x1* and *x2* and with the new edge *x1x2*. A modified *block_2* is a *block_2* where the edge *x1x2* is replaced by the path *x1—y—x2*; etc. When *block_n* page with its properties is added to the A-CGT-portal an observation is the following. Redefine the basic block thus by adding a new level of vertices edges: *new $V_r$ = old $V_r$ + {G, . . . , N}; new $E_r$ = old $E_r$ – {CD, CE, DF, EF} + {CG, CH, DI, DJ, EK, EL, FM, FN, GH, IJ, KL MN, GI, HJ, KM, LN}*. We have now defined a new family of *3*-regular graphs.

Ex. 2. Another problem in CGT involves the construction of a family of *4*-regular graphs, $G_k$, *k = 1, 2, 3, . . .* . To build $G_1$ we start with a square i.e., $C_4$ with vertices *A, B, C, D*. We next attach to these four vertices four new (vertex) labeled squares *AEFH, BHIJ, CKLM, DNPQ*. We add the additional edges *EG, HJ, KM, QN* in the outer squares. Finally we add the edges *EH, FI, JK, IL, MN, NP*. This completes the description of $G_1$. To build $G_2$ we start with a copy of $G_1$ with vertex labels as before. At the vertices *F, I, L, P* we attach new squares *FRST, IUVW, LXYZ, $PZ_1Z_2Z_3$*. We complete the construction of $G_2$ by adding the edges *RU, SV, WX, VY, $ZZ\_1$, $YZ_2$, $TZ_3$, $SZ_2$*. With $G_k$ defined the questions can be (a) With reasoning state how many $C_3, C_4, C_6$ are present. (b) Give the recurrence relations to determine the number of vertices and edges.

Note that the prerequisite box for the *block_n* page and for the $G_k$ page will be those embracing graph fundamentals including terms and phrases such as *k*-regular graphs, isomorphism, recurrence relations. A student contribution to *block_n* page or $G_k$ page can be a set of figures of the defined graph families. With examples 1 and 2 above a Computer Science student will be invited to explore more about regular graphs e.g., *Petersen graph, Harary graphs* and will look forward to lectures based on textbook-based theorems. A Group 1 student is likely to discover a general principle such as "duplicate copies of an arbitrary graph on *n* vertexes, selectively *vertex-joined* via new edges will increase appropriate vertex degrees by *1* giving a random graph on *2n* vertices" and will give an algorithm for the question "how to construct a *(k+1)*-regular graph from an *k*-regular graph".

Consider again the family *block_n* as an A-CGT-portal page and assume that the concept of distance in graphs is known to the students. An assignment exercise now will be: For the *block_n* find the sum of the distances between vertices *u* and *v*, the sum taken across all distinct *(u,v)* pairs. Further successor questions will be of the following types:

(a) For the graphs *block_n* and $G_k$ find the ratio of number of edges to number of vertices.
(b) Upto isomorphism, enumerate all the spanning trees of the Petersen graph and find their Wiener indexes.
(c) Get the definition of the Boolean hypercube of order *n*. (i) Code in *Java/C++* to compute the Wiener indexes for *n = 1* through *5*. (ii) Find a formula for the Wiener index of the *n*-hypercube.

It can be argued that the more-to-explore attribute and the prerequisite boxes will assist in the teaching/learning process. In the context of the above discussions, the following examples are illustrative:

(a) Under the more-to-explore page attribute in an appropriate A-CGT-portal page can be a short write-up culled from contemporary literature, to provide possible use of Wiener-type indexes in Biochemistry for identification of lead drug candidates. This will add a further impetus for a Group 1 Computer Science student to discover more about total distance and other concepts such as cliques/independent sets in Graph theory.
(b) It has been widely suggested (see [8] for example) that algorithm based problem-solving involves abstractions and it

is an essential component in software development, a major concern for Computer Science students. An added activity of algorithm development and/or coding, for Computer Science majors, via an A-CGT-portal is therefore desirable. To begin with, the more-to-explore page attribute in an A-CGT-portal will point to pertinent combinatorial algorithms and graph algorithms – these may be to previously seen lectures and will provide more references for Group 1/2 students. In addition, problems suitably guided by P2 prerequisite boxes are likely to suggest programming exercises e.g., computing the Wiener index, given the weighted adjacency matrix of a graph suggesting say breadth-first search or Floyd-Warshall algorithms. In this process one may expect that Group 1 students will also propose new algorithms and program codes.

## IV. Concluding Remarks

A remark is that summative assessment with regard to students' contributions is a difficult task in the A-CGT-portal context because collaborations and consultations of other sources are rather difficult to quantify. An immediate honorable mention in an A-CGT-portal is a partial solution but it may not bring out the excellence in Group 1 students. Direct questions in programming/traditional assignments and tests based on students' contributions are likely to yield unfavourable feedbacks from all other students when the competitive spirit is at a level beyond a requisite level which again is difficult to measure. A team of experienced moderators knowing the students' previous performance and extent of work is a solution approach.

An outcome of maintaining students' records in a log, on the activities of contributing to page attributes or contributions as new page on the A-CGT-portal is that they can be suitably guided to credit further offered electives such as Coding Theory, Topics in Graph theory based on their major, and do project works in later semesters subject to their interests and career/program goals. At the end of a CGT course a developed A-CGT-portal can be edited further to serve as a profitable learning resource for various interest groups.

The pedagogical value of an A-CGT-portal should be apparent. Its difference from an annotated set of Wiki-pages as a supplementary resource for unsupervised learning can also be appreciated. In a CGT course students' exploration of an A-CGT-portal will depend on attitudes, intellect and inquisitiveness. From the example illustrations it is possible to reason that an A-CGT-portal can judiciously contribute to the various levels of students' cognition as viewed from a psychological perspective, although a conclusion demands an experimental study and inference mechanisms. As pointed out in [8] the processes of abstracting, concretizing, synthesizing and analyzing are the mental faculties that are to be perhaps developed in teaching Computer Science – the extent to which an A-CGT-portal can contribute to these processes is an interesting question. Also, from the viewpoint of freshers and some Group 2 and Group 3 students, didactical questions need further study. We have not addressed many other important aspects such as design, formats, presentation, interfaces pertaining to an A-CGT-portal and its ownership and management, copyright issues, incorporation of students' feedbacks, security issues, in the context of portal development. However these are beyond the central theme in this paper. Knowledge representation in an A-CGT-portal is another pertinent factor that will contribute to the ease of learning concepts. It may be argued that the portal-idea suggested is more towards specializing in one associated subject rather than breadthwise knowledge acquisition across the wide spectrum of subjects in Computer Science. This is to be weighed against further dependent courses in the program and the stated program outcomes/ goals.

Acknowledgements

The author would like to thank Dr.S.Sundarrajan, National Institute of Technology, Tiruchirapalli for his supporting this work.

---

A136328    Wiener Index of the Odd graph

0, 3, 75, 1435, 25515, 436821, 7339332, 121782375, 2005392675, 32835436777,
535550923908, 8707954925033, 141270179732500, 2287544190032700, 36988236910737360,
597341791692978975, 9637351741503033075

| | |
|---|---|
| OFFSET | 1,2 |
| COMMENTS | The odd graph O_n (n>=2) is a graph whose vertices represent the (n-1)-subsets of {1,2,...,2n-1} and two vertices are connected if and only if they correspond to disjoint subsets. It is a distance regular graph. [Emeric Deutsch, Aug 20 2013] |
| REFERENCES | Kailasam Viswanathan Iyer, Some computational and graph theoretical aspects of Wiener index, Ph.D., dissertation, Dept. of Comp. Sci. & Engg., National Institute of Technology, Trichy, India, 2007. |
| LINKS | Table of n, a(n) for n=1..17. |
| | R. Balakrishnan, N. Sridharan and K. Viswanathan Iyer, The Wiener index of Odd graphs, J. Ind. Inst. Sci., vol. 86, no. 5, 2006. |
| FORMULA | A formula can be found in the Balakrishnan et al. reference. |
| | A formula is "hidden" in the 2nd Maple program. B(n) and C(n) are the intersection arrays of O_n, H(n) is the Hosoya-Wiener polynomial of O_n, and Wi(n) is the Wiener index of O_n. [Emeric Deutsch, Aug 20 2013] |
| EXAMPLE | a(2)=3 is the Wiener index of O_2 which is C_3. |
| | a(3)=75 is the Wiener index of O_3 which is the Petersen graph. |
| MAPLE | A136328d := proc(k) add( (2*j+1)*binomial(k-1, j)^2/(1+j), j=0..(k/2-1) ); |
| | %+2*add( (k-1-j)*binomial(k-1, j)^2/(1+j), j=floor(k/2)..(k-2) ); k*% ; end proc: |
| | A136328 := proc(n) binomial(2*n-1, n-1)*A136328d(n)/2 ; end proc: |
| | seq(A136328(n), n=1..20) ; |
| | #  R.J.Mathar, Sep 15 2010 |
| | B := proc (n) options operator, arrow: [seq(n-floor((1/2)*m), m = 1 .. n-1)] end proc: C := proc (n) options operator, arrow: [seq(ceil((1/2)*m), m = 1 .. n-1)] end proc: H := proc (n) options operator, arrow: (1/2)*binomial(2*n-1, n-1)*(sum((product(B(n)[r]/C(n)[r], r = 1 .. j))*t^j, j = 1 .. n-1)) end proc: Wi := proc (n) options operator, arrow: subs(t = 1, diff(H(n), t)) end proc: |
| | seq(Wi(n), n = 2 .. 20); |
| | # Emeric Deutsch, Aug 20 2013 |
| KEYWORD | nonn |
| AUTHOR | K.V.Iyer, Mar 27 2008 |
| EXTENSIONS | Extended by R.J.Mathar, Sep 15 2010 |
| STATUS | Approved |

---

Illustration 1: A sample page from the OEIS